\documentclass[prb,twocolumn,showkeys,showpacs,preprintnumbers,amsmath,amssymb]{revtex4}
\usepackage{graphicx}
\usepackage{dcolumn}
\usepackage{bm}
\usepackage[tight]{subfigure}
\usepackage{enumerate}
\usepackage{amsmath}
\usepackage{verbatim}
\usepackage{color}
\usepackage{natbib}

\begin{document}
\title{Electron-hole interactions in coupled InAs-GaSb quantum dots based on nanowire crystal phase templates}
\today
\author{Malin Nilsson$^{1}$}
\author{Luna Namazi$^{1}$}
\author{Sebastian Lehmann$^{1}$}
\author{Martin Leijnse$^{1}$}
\author{Kimberly A. Dick$^{1,2}$}
\author{Claes Thelander$^{1}$}

\affiliation{$^{1}$Division of Solid State Physics and NanoLund, Lund University, Box 118, S-221 00 Lund, Sweden}
\affiliation{$^{2}$Center for Analysis and Synthesis, Lund University, Box 124, S-221 00 Lund, Sweden} 
  
\keywords{Nanowire, crystal phase control, coupled quantum dot, InAs, GaSb, electron-hole interactions}  
  
\begin{abstract}
We report growth and characterization of a coupled quantum dot structure that utilizes nanowire templates for selective epitaxy of radial heterostructures. The starting point is a zinc blende InAs nanowire with thin segments of wurtzite structure. These segments have dual roles: they act as tunnel barriers for electron transport in the InAs core, and they also locally suppress growth of a GaSb shell, resulting in coaxial InAs-GaSb quantum dots with integrated electrical probes. The parallel quantum dot structure hosts spatially separated electrons and holes that interact due to the type-II broken gap of InAs-GaSb heterojunctions. The Coulomb blockade in the electron and hole transport is studied, and periodic interactions of electrons and holes are observed and can be reproduced by modeling. Distorted Coulomb diamonds indicate voltage-induced ground-state transitions, possibly a result of changes in the spatial distribution of holes in the thin GaSb shell. 
\end{abstract}

 \pacs{
73.23.Hk    
73.63.Kv 	
73.63.Nm 	
}
\maketitle

Coupled quantum dots (QDs) represent a model system for studies of charge carrier interactions, and are the heart of most schemes for spin-based quantum computation.\cite{Kloeffel2013} They can be realized by a number of methods, such as self-assembly of semiconductor nanocrystals\cite{Bjork2004,Stinaff2016} or electrostatic gating of a two-dimensional (2D) or one-dimensional (1D) semiconductor.\cite{Elzerman2003,Petta2005,Goss2013} Systems formed by electrostatic gating are generally studied by electrical measurements, whereas self-assembled QDs, which are difficult to access electrically, are almost exclusively investigated by optical techniques.
Coupled QDs are typically of the same material due to the complexity involved in the fabrication. However, use of dissimilar materials could provide for extended opportunities in, for instance, manipulation of strain,\cite{Christodoulou2015} $g$-factors,\cite{Bjork2005} and properties of spatially separated excitons.\cite{deMello2010} At present, such heterogeneous QDs are produced primarily through colloidal chemistry in the form of core-shell nanocrystals, and they are studied using optical methods. Here, applications often involve type-I band alignments, with a wider band gap shell for passivation purposes. However, type-II staggered band alignments are also interesting because they allow for tuning of the effective crystal band gap for emission below the core and shell band gaps.\cite{Kim2003}

Junctions of GaSb-InAs constitute a type-II band alignment system, which is of particular interest due to a broken gap, where valence- and conduction-band states overlap in energy. Experiments on tunable InAs-GaSb 2D junctions show that a hybridization gap can form in the 2D interior, while gapless topologically protected 1D states appear on the edges.\cite{Knez2014,Qu2015}  Other interest in the junction stems from the prediction of an excitonic ground state, with spatially separated electrons and holes.\cite{Naveh1996} Recently, low-temperature transport experiments in coupled GaSb-InAs QDs based on GaSb-InAs core-shell nanowires were reported.\cite{Ganjipour2015} For such nanowires, a coupled QD may form as a consequence of source and drain contact formation, but with limited control of tunnel barriers and confinement. The measurements showed, however, the coexistence, and interactions, of electrons and holes.

In this work, we investigate electron and hole transport in a coupled, InAs-GaSb core-shell QD structure formed entirely by epitaxy. The highly controlled self-assembled process allows for the parallel formation of an arbitrary number of structures, designed with built-in tunnel junctions, and with no requirement of advanced lithography and etching. The coupled dots formed have a core of InAs, and a shell of GaSb. They are obtained by controlling the InAs crystal structure along the growth direction of a nanowire, where the nanowire core subsequently acts as a 1D template for determining its final  structure,\cite{Namazi2015} see Fig.~\ref{fig1}. More specifically, segments of wurtzite (WZ) InAs in an otherwise zinc blende (ZB) nanowire locally suppress GaSb shell growth due to their relative low surface energy. A second role of the WZ segments is to act as tunnel barriers\cite{Dick2010,Nilsson2016}, thus allowing for control of the electron and hole populations in the parallel-coupled QDs by Coulomb blockade. 

We use single-electron tunneling transport in the InAs core to monitor single-hole charging events in the GaSb shell. In contrast to reference nanowires without a GaSb shell, exhibiting regular Coulomb oscillations, we find an obvious beating pattern in the Coulomb blockade pattern for the coupled InAs-GaSb QDs due to transport of spatially separated electrons and holes. We use complementary high-resolution scanning electron microscope (SEM) imaging on the measured devices, together with transmission electron microscope (TEM) imaging of reference structures, to estimate the QD dimensions, and we can accurately model the electrostatic electron-hole interactions in a number of structures. Furthermore, we observe Coulomb blockade properties that go beyond the simple picture of the constant interaction model, properties that are not observed for reference, core-only QDs. 

\section{Method}
The InAs-GaSb core-shell QDs are grown by means of metal organic vapor phase epitaxy (MOVPE) on InAs ($\bar{1} \bar{1} \bar{1}$) oriented InAs substrates with pre-deposited aerosol Au seed particles. In this study, seed particles of 30~nm diameter with an areal density of 1 particle per $\mu$m$^{2}$ were used. A 3x2'' close-coupled showerhead reactor from Aixtron with a total flow of 8~l/min and a reactor pressure of 100~mbar was used. The carrier gas was H$_{2}$, with trimethylindium (TMIn) and arsine (AsH$_{3}$) as precursors for InAs, and trimethylgallium (TMGa) and trimethylantimony (TMSb) for GaSb. Prior to growth, the samples were annealed with H$_{2}$/AsH$_{3}$ (molar fraction of AsH$_{3}$ = 2.5$\times$10$^{-3}$ at a set temperature of 550$^{\circ}$C) to remove native oxides from the surface of the substrates. The growth temperature for both the InAs core and the GaSb shell was set to 470$^{\circ}$C. The crystal structure of the InAs core was controlled by tuning the V/III precursor ratio during the growth process; a higher V/III ratio results in ZB crystal structure.\cite{Lehmann2013} The molar fraction of AsH$_{3}$ was set to 2.3$\times$10$^{-5}$, and 2.5$\times$10$^{-2}$, for WZ and ZB respectively, while TMIn was kept at 1.8$\times$10$^{-6}$, for both crystal phases. The TMGa and TMSb flows for the GaSb shell were both set to 2.7$\times$10$^{-5}$.  

Two sets of samples were studied: an InAs reference sample and an InAs-GaSb core-shell sample, both with a growth time of 11~s for the WZ segments and 30~s for the enclosed ZB segment. The GaSb shell was grown for 20 minutes by switching off the TMIn and AsH$_{3}$, and switching on the TMGa and TMSb precursors simultaneously. Also, in order to passivate the GaSb surface,\cite{Ganjipour2014} an additional thin shell of InAs was grown for 3 minutes. The growth parameters for the outer InAs shell were set to the same values used for growing the WZ segments of the core. For more details on the growth of core-shell InAs-GaSb nanowires, see Namazi \textit{et al.}\cite{Namazi2015}

Nanowires were transferred mechanically onto lacey carbon covered copper grids for structural and compositional TEM analysis in a JEOL 3000F. Conventional dark-field imaging along a $\langle 21\bar{3}\rangle$-type axis using the $000\bar{1}$ double diffraction spot was employed for crystal-structure-sensitive TEM imaging. Radial shell thicknesses in turn were extracted from energy dispersive X-ray spectroscopy (XEDS) collected in scanning TEM high-angle annular dark-field (HAADF) imaging mode along a $\langle 11\bar{2}\rangle$-type direction. For acquisition of SEM data of fully processed nanowire devices, a Zeiss Leo Gemini 1560 setup operated at 20~kV and beam currents on the order of 100 $\mu$A were used. To ensure satisfying conditions for electron channeling contrast imaging (ECCI), and thus to distinguish WZ and ZB segments in the nanowires, sample tilts in the range of -5$^{\circ}$ - 20$^{\circ}$ were set.\cite{Nilsson2016,Zaefferer2014,Joy1982} 

To perform electrical characterization of the QDs, the nanowires were mechanically transferred from the growth samples to degenerately $n$-doped silicon substrates with a 110~nm thick SiO$_{2}$ film. The back sides of the substrates are covered with gold and the doped Si substrate serve as a global back gate during the electrical measurements. Predefined contact pads and coordinate systems on the top sides of the substrates facilitate aligning of source and drain contacts on suitable nanowires selected by SEM imaging. Source and drain contacts are fabricated by means of electron beam lithography and a standard lift-off process.  Prior to metallization, resist residues and native oxides were removed by a 30~s O$_{2}$-plasma etch and an HCl:H$_{2}$O (1:20) etch for 10~s, respectively. The contacts consist of 25~nm Ni and 75~nm Au, and they are separated by 950~nm.

\section{Results and Discussion}

In bulk, the ZB InAs-GaSb heterojunction exhibits a type-II broken band alignment where the InAs conduction band and the GaSb valence band overlap by approximately 150~meV. In a quantum confined system, the overlap of corresponding electron and hole states can be tuned by the confinement energy.\cite{Kishore2012,Ganjipour2012} 
\begin{figure}[b]
\centering
\includegraphics[width=\columnwidth]{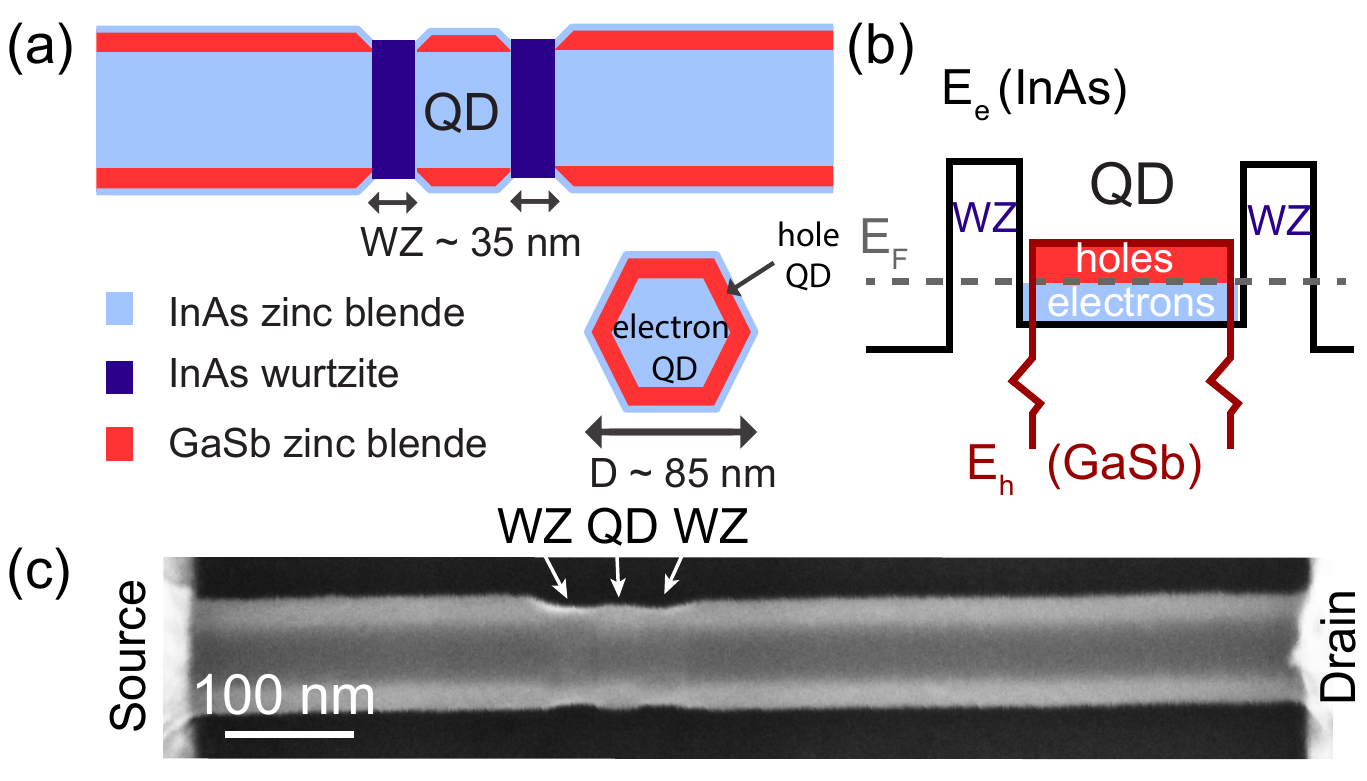}
\caption{(Color online) (a) Schematic illustrations of the core-shell QD structure where the hexagonal cross section is indicated. (b) A simplified energy band diagram of the double QD, where dark red corresponds to the lowest hole energy states $E_h$ in the GaSb valence band, and black corresponds to the lowest electron energy states $E_e$ in the InAs conduction band. (c) A SEM image of a core-shell QD NW device, where the WZ segments and the QD are indicated. }
\label{fig1}
\end{figure}
Schematic illustrations of the core-shell QD structure are presented in Fig.~\ref{fig1}(a), showing average values for WZ segment lengths (35 nm) and diameter (85 nm). The WZ segments in the InAs act as tunnel barriers for electron transport, effectively defining a QD\cite{Nilsson2016}, and due to the suppression of GaSb radial growth, they also provide barriers for hole transport in the GaSb shell.

 A simplified band diagram of the core-shell QD is displayed in Fig.~\ref{fig1}(b), where black lines represent the lowest electron energy states in the InAs core conduction band, and dark red lines represent the lowest hole energy states in the GaSb shell valence band.  Quantum confinement, particularly in the very thin GaSb shell, reduces the energy overlap of electron and hole states of the two materials. The thin, outer passivating InAs shell is omitted here due to a strong confinement-induced shift of these energy levels.\cite{Ganjipour2012}  Devices with no such outer InAs shell exhibit poor hole transport characteristics. Figure~\ref{fig1}(c) shows an SEM image of a core-shell QD NW device, where the positions of the two WZ segments have been indicated.

\begin{figure}
\centering
\includegraphics[width=0.8\columnwidth]{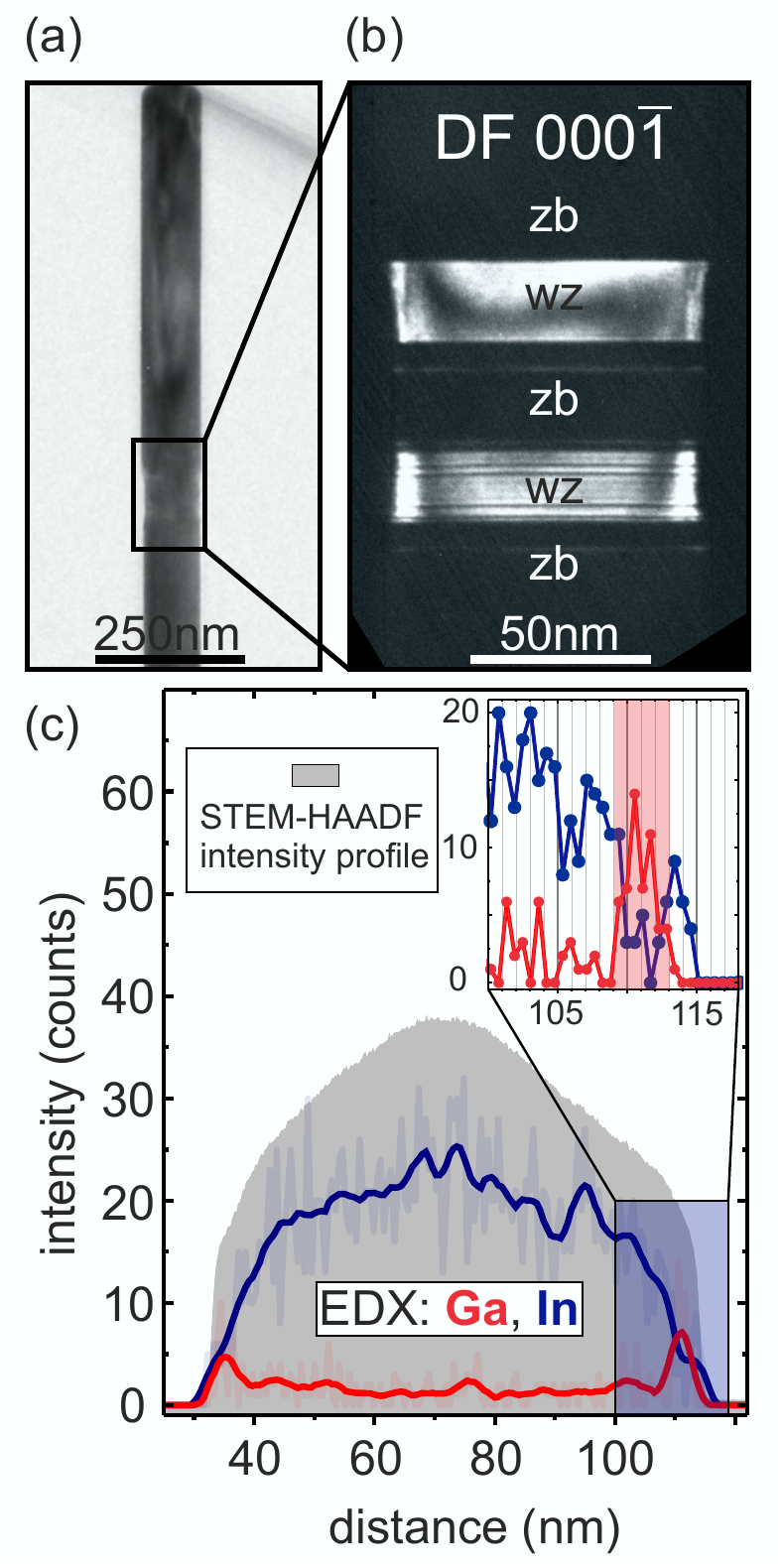}
\caption{(Color online) (a) Bright-field TEM image of a nanowire containing a core-shell QD viewed along a $\langle 11\bar{2}\rangle$-type direction (b) Dark-field TEM image recorded along a $\langle21\bar{3}\rangle$-type direction and selecting the WZ 000$\bar{1}$ double diffraction spot for imaging, where WZ appears as bright contrast and ZB as dark contrast. (c) EDX line scan recorded across the ZB QD perpendicular to the nanowire growth axis, averaged over a width of 20 lines and showing the Ga K$\alpha$ (red) and In L$\alpha$ (blue) signal. The inset is a magnification and highlights the signals from the GaSb and InAs shells. The STEM-HAADF intensity profile is also given (gray background), which reflects the morphology of the nanowire at that position. }
\label{fig2}
\end{figure}

A bright-field overview TEM image of a typical core-shell nanowire is displayed in Fig.~\ref{fig2}(a). In the high resolution dark-field image shown in Fig.~\ref{fig2}(b), the WZ segments (19/22~nm) defining the QD (31~nm) are clearly visible as bright contrast. Additional bright lines in the ZB segments are signals from twinned segments in the ZB crystal phase. However, we do not find any impact from such twinned segments on the electrical characteristics.\cite{Nilsson2016} From the compositional analysis in the radial direction on the QD segment displayed in Fig.~\ref{fig2}(c), we find a GaSb shell thickness of 3-5~nm. The analysis also shows the presence of the even thinner, outer passivating InAs shell, with a thickness of 1-3~nm (hence thin enough not to contribute to electron transport due to strong confinement).\cite{Ganjipour2012}

\begin{figure*}
\centering
\includegraphics[width=\textwidth]{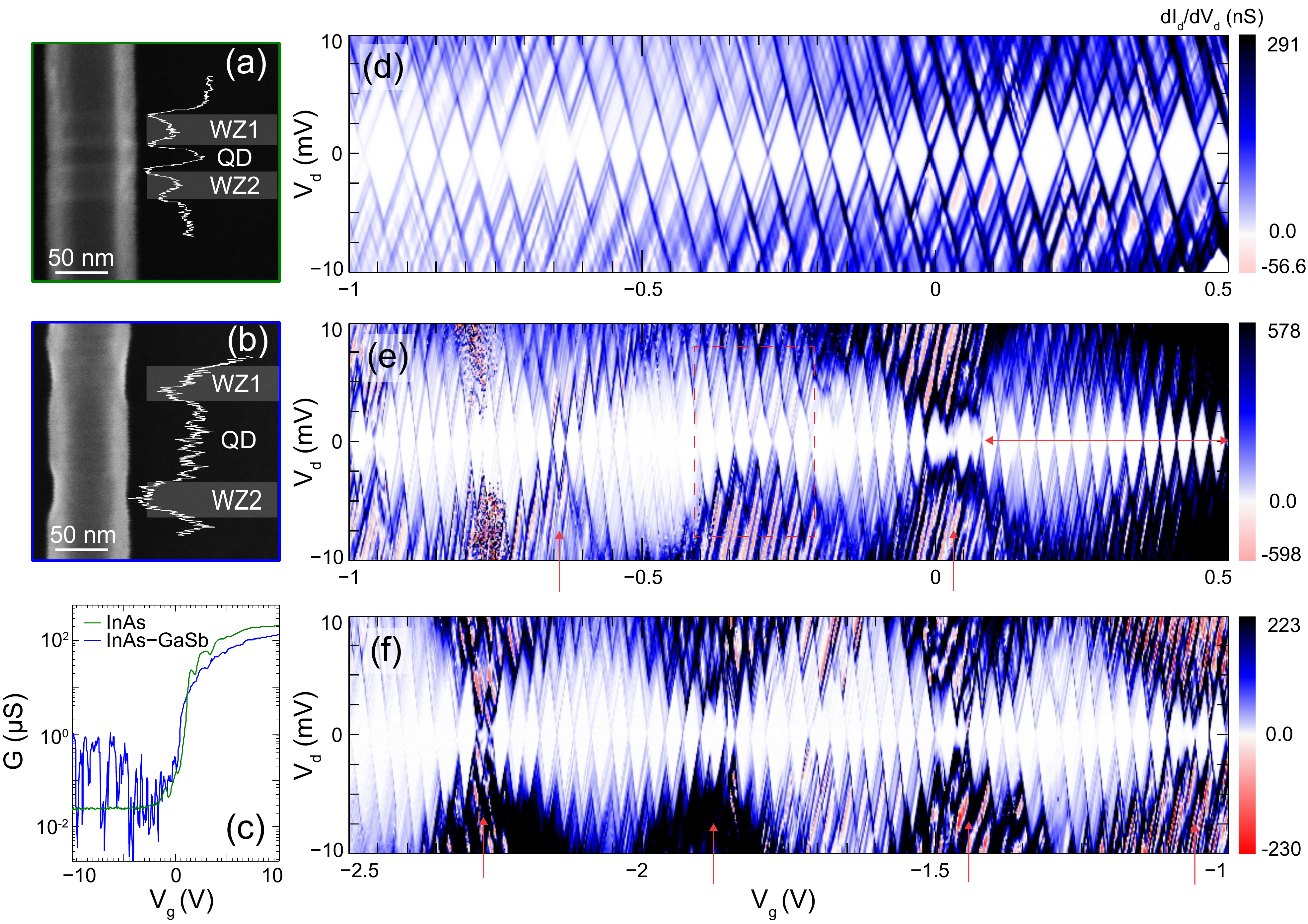}
\caption{(Color online) SEM images together with intensity profiles of measured QDs; (a) reference bare InAs ("WZ1" = 29 nm, "QD" = 24 nm, "WZ2" = 24 nm) and (b) core-shell InAs-GaSb ("WZ1" = 29 nm, "QD" = 73 nm, "WZ2" = 28 nm). The diameters of the NWs at the QD segment are 84 and 76~nm, respectively. The accuracy of all given dimensions is $\pm$2~nm. (c) Conductance as a function of gate voltage at a bias of 10 mV recorded for the reference QD (green) and core-shell QD (blue). Charge stability diagrams recorded for: (d) the reference QD, and (e) and (f) the core-shell QD. The vertical red arrows indicate lifting of the hole blockade.  }
\label{fig3}
\end{figure*}

Next, the electrical properties of the QD structure are investigated at an estimated electron temperature of 400~mK. Figures~\ref{fig3}(a) and \ref{fig3}(b) show SEM images of a bare reference InAs QD and a core-shell QD, respectively, from which WZ tunnel barrier and QD dimensions have been extracted. The corresponding electrical transport data are shown in Figs.~\ref{fig3}(c)-(f). Starting with the conductance trace as a function of back-gate voltage [Fig.~\ref{fig3}(c)], we observe that the InAs reference nanowire (green) becomes depleted for negative gate voltages, whereas the core-shell nanowire (blue) shows ambipolar behavior as a result of an onset of hole transport in the GaSb shell for $V_g <$ 0~V.\cite{Gluschke2015} Charge stability diagrams were recorded for both devices, and they are shown in Figs.~\ref{fig3}(d)-(f). The reference InAs QD [Fig.~\ref{fig3}(d)] exhibits clear and regular Coulomb blockade diamonds with a gate capacitance $C_g$ of 3.3 ($\pm$0.5)~aF and a charging energy $E_C$ of approximately 3.3 ($\pm$0.5)~meV, consistent with earlier reports.\cite{Nilsson2016} Near $V_g = 0$~V, a pattern of one large and three small diamonds, consistent with fourfold degeneracy, is visible. This is expected for a symmetric cylindrical system,\cite{Bjork2004,Ford2012} indicating low defect density on the surface of the nanowire. At other gate voltages a non-uniform vertical potential lifts the symmetry, and the fourfold degeneracy disappears. 

In contrast, the core-shell QD [Figs.~\ref{fig3}(e) and \ref{fig3}(f)] shows a pronounced composite pattern of large Coulomb diamonds, overlaid with small, regular diamonds. This pattern is typical for a system of two QDs coupled in parallel to the source and drain contacts. Taking into account the relatively longer core-shell QD length, as evident from the SEM images, we expect a correspondingly smaller $\Delta V_g$~$(= q/C_g)$, and thus smaller diamonds, compared to the reference InAs device. We can therefore conclude that the small diamonds in Figs.~\ref{fig3}(e) and \ref{fig3}(f) are due to single-electron charging of the InAs core QD. This is also consistent with an increasing conductance of the Coulomb oscillations associated with these diamonds for increasing positive $V_g$. 

We note that for positive $V_g$ in Fig.~\ref{fig3}(e) there is a regime of unperturbed electron diamonds $(\Delta V_g \approx 30$~$(\pm1)$~mV, $C_g \approx 5.5$~$(\pm1)$~aF), indicating a stable charge in the vicinity of the core QD and a sign of a possibly depleted shell.  However, for negative gate voltages, the regular blockade pattern is disturbed at specific intervals due to changes in the charge outside of the core QD. Here, a superimposed pattern of larger diamonds emerges, which is consistent with an onset of hole transport via a smaller QD, formed in the GaSb shell. Within a large diamond the overall conductance is suppressed, as is expected for a parallel QD system, and here the small Coulomb diamonds are generally unperturbed, with a similar $\Delta V_g$~$(\approx 30$~mV) as noted before. In Figs.~\ref{fig3}(f) and \ref{fig3}(e), clear transitions between hole diamonds are marked with red horizontal arrows. However, in the region indicated with a dashed rectangle in Fig.~\ref{fig3}(e), a distinct transition seems to be missing, where a series of Coulomb diamonds instead show slopes with pronounced kinks. Similar features have been reported for single carbon nanotubes by Tans \textit{et al.}\cite{Tans1998}, where the authors attribute the kinks to internal transitions due to electron-electron interactions.

 For this sample, we have followed the beating pattern to a gate voltage of -11~V, at which point the population of the electron QD can still be observed; see Fig.~\ref{fig6} in the Appendix. Generally, we expect that electron-hole QDs are more difficult to deplete of either charge carrier due to the attractive electrostatic interactions that act to screen the gate potential.

\begin{figure*}
\centering
\includegraphics[width=\textwidth]{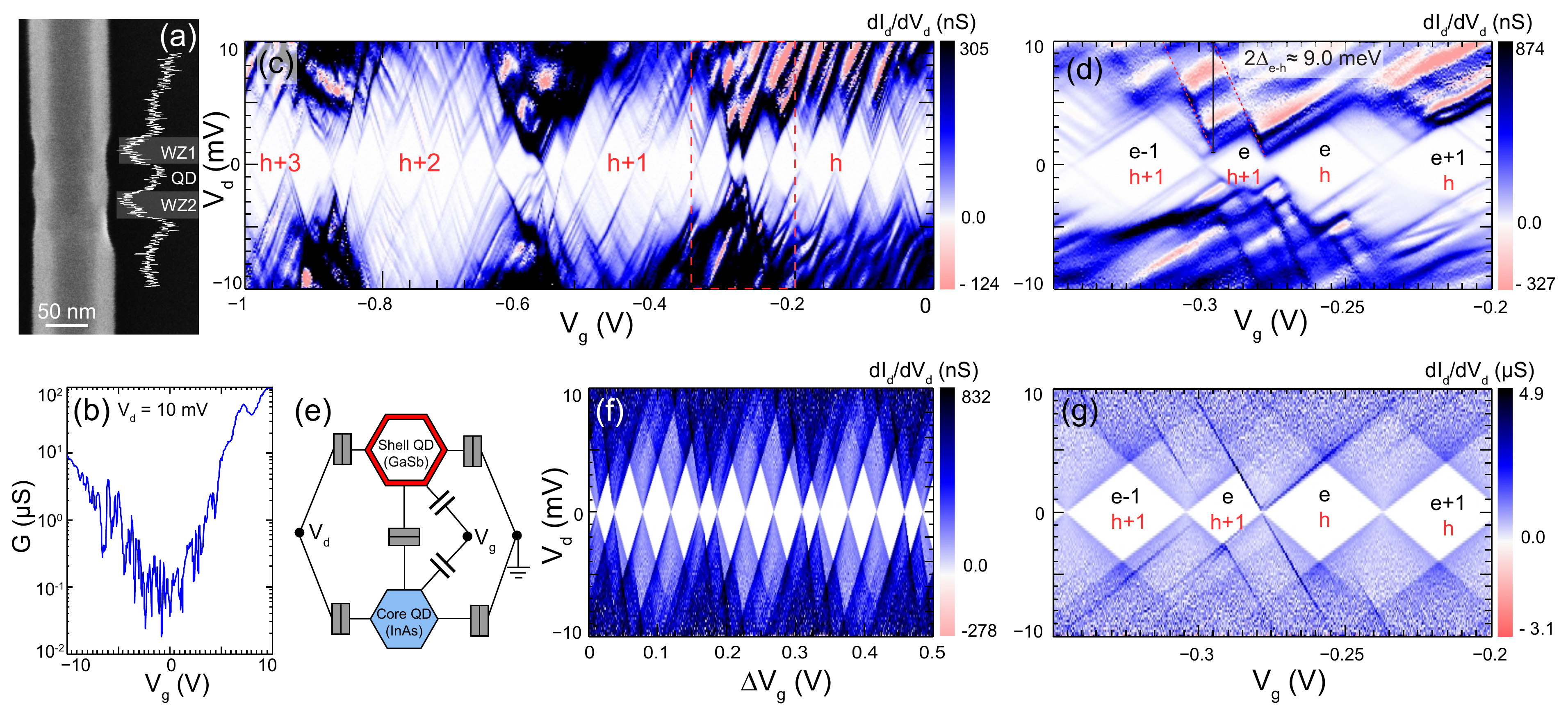}
\caption{(Color online) (a) SEM image of a second core-shell QD device including an intensity profile ("WZ1" = 36 nm, "QD" = 37 nm, "WZ2" = 36 nm). The diameter of the segment containing the QDs is 91 nm. The accuracy of all given dimensions is $\pm$2 nm (b) Conductance as a function of gate voltage.  (c) Measured charge stability diagram, where the larger diamonds corresponding to hole blockade are marked and where the dashed rectangle indicates the area displayed in panel (d). (d) A close-up of a transition between two hole-Coulomb-blockade regions. The black and red notations correspond to a suggested electron and hole occupation, respectively. An electron-hole interaction strength $\Delta_{e-h}$ of approximately 4.5~meV is extracted from the shift in the onset of hole conduction. (e) Equivalent circuit schematic of the system used in the modeling. (f) and (g) Simulated charge stability diagrams of a coupled QD system with two interacting islands connected to a source and drain electrode. The values for all capacitances are the same in the two simulations. (f) Both QDs modeled as metallic islands with no confinement (gate voltage span of 0.5~V). (g) Fit to panel (d), based on a system of one metallic island and a QD with a single level only.  }
\label{fig4}
\end{figure*}

We next investigate electrostatic interactions between electrons and holes, where we focus on a second core-shell QD device, as shown in Fig.~\ref{fig4}(a). The $G(V_g)$ measurement in Fig.~\ref{fig4}(b) for this device reveals clear ambipolar characteristics, with an apparent transition from electron- to hole-dominated transport near $V_g = 0$~V. A corresponding charge stability diagram is shown in Fig.~\ref{fig4}(c) where the previously discussed beating pattern is clearly visible. A notable difference from the device in Figs.~\ref{fig3}(e) and \ref{fig3}(f) is, however, that fewer small diamonds fit inside each large diamond for this coupled QD. The small diamonds, which we attribute to charging of the InAs core, here have approximately a factor of 2 smaller gate coupling $\alpha \sim \Delta V_d/\Delta V_g$  (0.080 versus 0.13), which agrees with the factor of 2 shorter QD length (37~nm versus 73~nm) as determined by SEM.

A suggested hole occupation in the stability diagram is marked with red symbols in Fig.~\ref{fig4}(c). At points where the hole blockade is lifted, such as the crossing from $h+1$ to $h$ (red dashed rectangle), we see clear effects of electrostatic interactions between electrons and holes. Figure~\ref{fig4}(d) shows a magnification of this hole-crossing, where electron and hole occupations are indicated in black and red text, respectively. Coming from the left, the peak in the $dI/dV_d$ indicated with a dashed red line (with strong gate coupling), corresponds to the removal of a hole. This line shifts to the right when an electron is added to the system. We attribute this shift to an attractive interaction between spatially separated electrons and holes in the core and shell, and we extract an electron-hole interaction strength $\Delta_{e-h}$ of 4.5~meV, which is an upper estimate assuming symmetric capacitive coupling to source and drain. Similar behavior has been reported for the reverse (GaSb-InAs) core-shell system.\cite{Ganjipour2015} Also in scanning tunneling microscopy of type-II staggered PbSe-CdSe core-shell colloids, it was shown that for some biasing conditions, electron tunneling was only possible when a hole was present in the core of the colloid.\cite{Swart2010}

Numerical Monte Carlo simulations of the parallel coupled QD system were performed using SIMON,\cite{Wasshuber2001} where capacitances and tunnel resistances were used as input parameters, with a temperature set to 0~K. A schematic of the equivalent circuit of the system used in the numerical simulations is displayed in Fig.~\ref{fig4}(e). Results from the simulations are presented in Figs.~\ref{fig4}(f) and \ref{fig4}(g). In Fig.~\ref{fig4}(f), a system of two metal dots was used, where the fit replicates the measurement well [Fig.~\ref{fig4}(c)], with the exception of one less small diamond per large diamond. However, the model does not take into account quantum confinement. In Fig.~\ref{fig4}(g), a fit to Fig.~\ref{fig4}(d) was made using one metallic island interacting with a second QD having a single level only. 
\begin{figure*}[t]
\centering
\includegraphics[width=\textwidth]{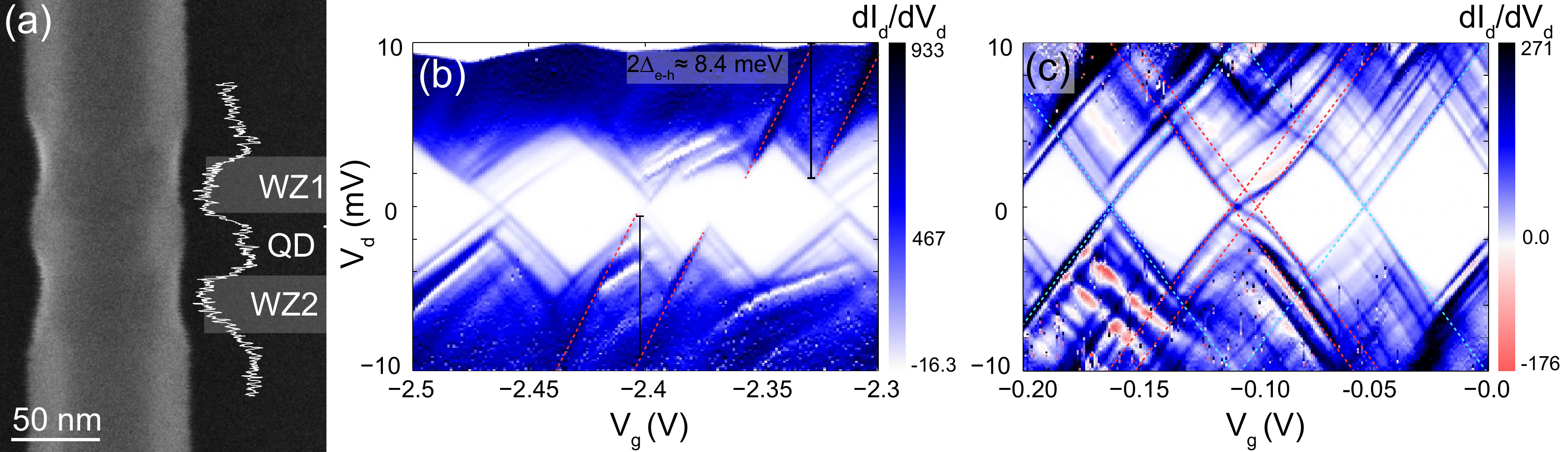}
\caption{(Color online) (a) SEM image of a third core-shell QD device including an intensity profile ("WZ1" = 31 nm, "QD" = 33 nm, "WZ2" = 30 nm). The diameter of the segment containing the QDs is 83~nm. The accuracy of all given dimensions is $\pm$2~nm. (b) and (c)  Charge stability diagrams for two different $V_g$ intervals for the corresponding device. In (b), $\Delta_{e-h}$ of approximately 4.2~meV is extracted from the shift in the onset of hole conduction (red dashed lines).  In (c), dashed blue and red lines indicate the expected borders of unperturbed Coulomb diamonds.}
\label{fig5}
\end{figure*}
With this model it is possible to reproduce most details in the experiment that involve ground-state transport. Details of all capacitances and resistances in the fit are shown in the Appendix.

In addition to the distinct shifts in the electron diamond pattern caused by changes in the hole population, some devices also exhibit another type of change in charge configuration as sensed by the core QD. Figures~\ref{fig5}(a)-\ref{fig5}(c) show results, including a close-up SEM image, from a third core-shell QD device that exemplifies this behavior. In line with previous observations, Fig.~\ref{fig5}(b) shows an example of a distinct shift in a hole conduction line with an extracted $\Delta_{e-h}$ of 4.2~meV. For other gate-voltage regions, such as that shown in Fig.~\ref{fig5}(c), we note a second type of charge transition that is more gradual, as if the gate potential is partially screened. Notably, the total electrostatic shift is weaker than for the case in Fig.~\ref{fig5}(b). This could, for instance, be explained by interactions with a smoothly changing population of states outside of the QD. It could also be a sign of charge reconfiguration in the shell, such that rather than a change in hole population, the spatial  distribution of holes in the shell changes with gate voltage. 


\section{Summary}
A coupled electron-hole QD system was realized using nanowire crystal phase templates that spatially determine radial heterostructure growth. It was found that pairs of InAs WZ crystal phase segments effectively inhibit GaSb radial growth, and provide tunnel barriers to an enclosed coupled QD. Low-temperature measurements of such parallel QDs reveal an overlay of Coulomb diamonds originating from hole transport in the GaSb shell with corresponding diamonds from electron transport in the InAs core. In addition to clear shifts in the electron energies with a change in hole population, we also observe curved and kinked Coulomb diamonds, which may indicate screening, potentially from a changing spatial distribution of holes in the thin GaSb shell.

The scalable fabrication concept used in this work should enable future studies on even smaller, and more complex, interacting systems. The hole QDs studied here are likely in the few-hole regime, whereas the electron QDs have many electrons. However, the few-electron regime should be reachable by increasing the electron confinement through a reduced ZB segment length in the epitaxial growth process.

\begin{acknowledgments}
The authors thank A. Burke for support with the electrical measurement setup. This work was carried out with financial support from NanoLund, the Swedish Research Council (VR), the Swedish Foundation for Strategic Research (SSF), and the Knut and Alice Wallenberg Foundation (KAW).
\end{acknowledgments}

\appendix*
\section{•}
\begin{figure*}
\centering
\includegraphics[width=\textwidth]{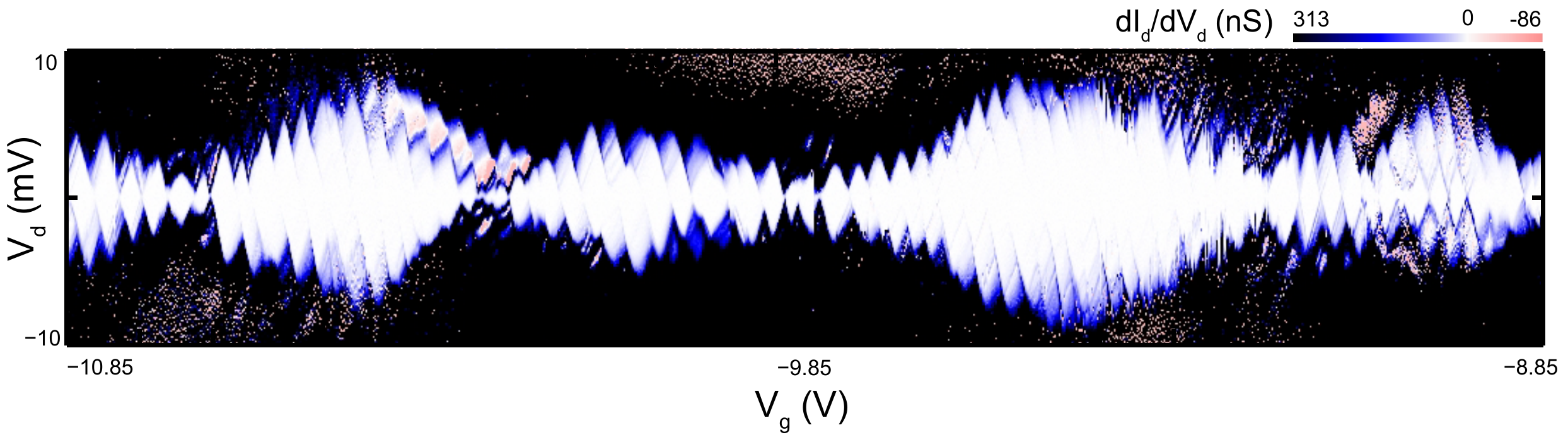}
\caption{(Color online) Charge stability diagram recorded from the core-shell device discussed in Fig.~\ref{fig3}. }
\label{fig6}
\end{figure*}

Figure~\ref{fig6} shows a charge stability diagram recorded at larger negative $V_g$ well beyond where the electron transport in typical reference InAs QDs is fully suppressed. However, in the case of core-shell devices, the holes in the GaSb shell partially screen the gate potential, leading to a continuation of electron transport even at large negative $V_g$.

\begin{figure}
\centering
\includegraphics[width=0.8\columnwidth]{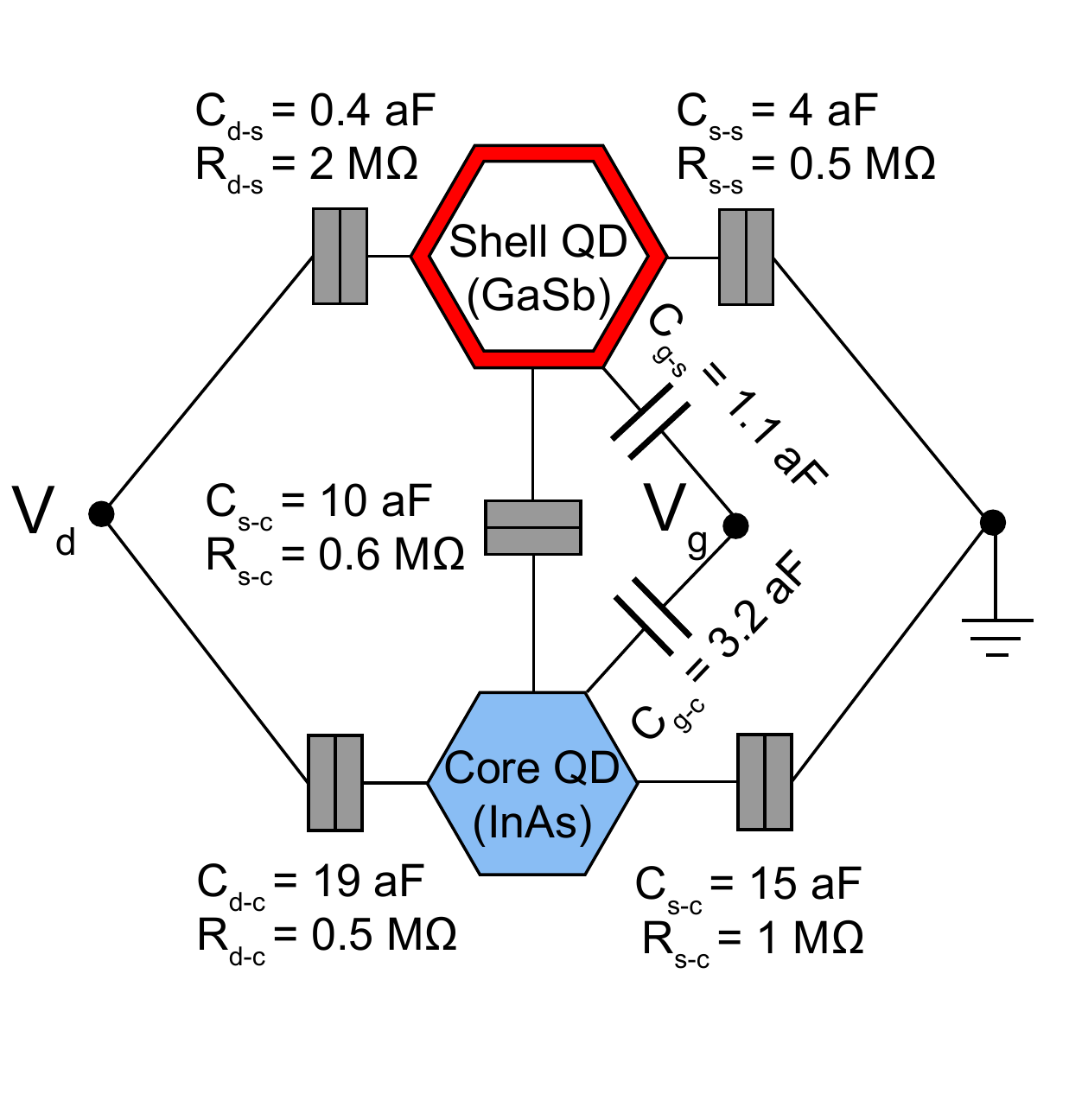}
\caption{(Color online) Equivalent circuit of the parallel coupled QD system, with the values for capacitances and resistances that were used in the simulation displayed in Fig.~\ref{fig4}(g). }
\label{fig7}
\end{figure}

Figure~\ref{fig7}  shows an equivalent circuit of the parallel-coupled QD with values for capacitances and resistances used in the simulation displayed in Fig.~\ref{fig4}(g). The same capacitance values were used in the simulation displayed in Fig.~\ref{fig4}(f). The fit was  performed according to the following scheme: (I) A single  QD (1) is first inserted that replicates the charging energy, Coulomb oscillation frequency, and tunnel resistances of the small Coulomb diamonds (attributed to transport in the InAs core).
(II) A second QD (2) is added to the system, focusing first on tuning the gate capacitances of the two dots, and the intra-QD capacitance to replicate the observed shift in the Coulomb diamond pattern related to QD 1 as QD 2 is charged. 
(III) Source-drain capacitances and resistances of QD 2 are adjusted to replicate the slopes and strengths of the Coulomb blockade borders of QD 2.
(IV) An offset charge is added to QD 2 to replicate the crossing point of the Coulomb blockade patterns.
(V) All capacitances are then slightly adjusted iteratively to improve the fit, and also the offset charge is adjusted. This is necessary since the QDs are coupled; introducing QD 2 affects the capacitances of QD 1. 
(VI) Finally, the fit is assessed by studying the Coulomb blockade pattern on a larger $V_g$ scale, where multiple Coulomb diamonds related to QD 2 can be observed. We check that the gate coupling to this dot is within reason, and that the slopes of the Coulomb diamond borders are replicated.

\bibliography{References_CS}

\end{document}